\documentclass[review]{elsarticle}

\usepackage{lineno}
\usepackage{amsmath}
\usepackage{amssymb}
\usepackage{ragged2e}
\usepackage{color}
\journal{Chaos, Solitons and Fractals}

\begin{document}


\begin{frontmatter}
\title{Tunable beam splitting via photorefractive nonlinearity and its applications in chiral waveguide induction and vortex generation}


\author[label1,label2]{Hechong Chen$\dagger$}
\author[label1,label2]{Zihan Liu$\dagger$}
\author[label1,label2]{Shengdi Lian}
\author[label1,label2]{Qingying Quan}
\author[label3]{Boris A. Malomed}
\author[label4]{Shuobo Li}
\author[label1,label2]{Yong Zhang}
\author[label5]{Huagang Li\corref{mycorrespondingauthor}}
\author[label1,label2]{Dongmei Deng\corref{mycorrespondingauthor}}

\cortext[mycorrespondingauthor]{Corresponding author}
\ead{lhg\_3@sina.com and dmdeng@263.net}

\affiliation[label1]{organization={School of Information and Optoelectronic Science and Engineering, South China Normal University, Guangzhou 510006, China}
               }
\affiliation[label2]{organization={Guangdong Provincial Key Laboratory of Nanophotonic Functional Materials and Devices, South China Normal University, Guangzhou 510631, China}
               }
\affiliation[label3]{organization={Department of Physical Electronics, School of Electrical Engineering, Faculty of Engineering, Tel Aviv University, Tel Aviv 69978, Israel}
               }
\affiliation[label4]{organization={School of Science, Sun Yat-sen University, Guangzhou 510275, China}
               }
\affiliation[label5]{organization={School of Photoelectric Engineering, Guangdong Polytechnic Normal University, Guangzhou 510665, China}
               }

\begin{abstract}
We report experimental observation and theoretical explanation of novel propagation regimes for optical beams in an artificial nonlinear material with outstanding photorefractive properties. Nondiffractive beams, which keep their shapes invariant in the free space, feature self-splitting from the middle in two separating secondary beams, due to the light-matter interaction. The splitting degree is controlled by means of a phase-pre-modulation method. We propose an application of the self-splitting to the creation of an effectively chiral waveguide and the generation of even-order vortices.
\end{abstract}

\end{frontmatter}


\section{Introduction}

\noindent As the development of the integrated-circuit technology is driving
the Moore's law towards its limits \cite{1}, the demand for new technology
platforms is growing. Photonic devices offer superior characteristics for
the design of circuitry, such as low loss, high modulation speeds, the
ability to implement multi-channel schemes, and others \cite{2,3,4}.
Material elements used for the construction of advanced photonic setups
include photonic crystals \cite{5,6}, metamaterials \cite{7,8},
quantum-limitation materials \cite{9,10}, electro-optic crystals \cite{11},
etc. Thus, a relevant direction of experimental and theoretical work is the
study of light-matter interactions in these media, which reveal phenomena
such as topological photonics \cite{12,13,14}, the Kerr, Pockels, and
higher-order nonlinear effects \cite{15,16}, propagation of structured light
in metamaterials \cite{17}, etc.

In particular, the nonlinearity represented by the photorefractive effect
has drawn much interest. It was first observed in 1966 by Ashin \textit{et al%
}. \cite{18} as wave-front distortions in a coherent light beam propagating
through a Lithium Niobate crystal. Governed by the nonlinear Schr\"{o}dinger
equation \cite{19}, coupled to the potential equation derived from the
band-transport model proposed by Kukhtarev \textit{et al}. \cite{20},
photorefractive crystals offer a highly efficient platform for light
modulation, in the reciprocal \cite{21,22} and real \cite{23,24,25,26}
spaces alike. While the model of the photorefractive effect is a
well-established one, novel phenomena in this area have been recently
demonstrated. In particular, Ma \textit{et al.} observed the effective
breakdown of the Newton's third law in photorefractive materials
such as the
strontium barium niobate (SBN) crystal \cite{27}, while Xia \textit{et al}.
investigated topological phenomena on this platform \cite{22,28}, Diebel
\textit{et al}. produced sinusoidal-shaped spatial solitons \cite{29}, and
Armijo \textit{et al}. and Schwartz \textit{et al}. studied
light propagation in disordered photonic lattices with the photorefractive
nonlinearity \cite{30,31}.

Nondiffractive beams, featuring invariance of the energy
distribution in the course of the propagation, are geometrically protected
by the caustics \cite{32,33}. Due to their unique robustness and longer
Rayleigh range, compared to the traditional Gaussian beams, nondiffractive
ones have been proposed for the use in complex-environment communications
\cite{34} and laser processing \cite{35}. Recently, many works have used
nondiffractive beams in SBN crystals, with the aim to confine light in the
course of the long-distance propagation \cite{36}. In Ref. \cite{37}, the research of analogous optical superfluid, Bessel
beam was used to create an obstacle of a few microns diameter, persisting along the propagation distance measured in several centimeters.

While the diffraction-free propagation is defined as a linear property, the study of effects of optical nonlinearities on these propagation regimes is a relevant problem too. In this article, we address the nonlinear propagation of nondiffractive beams in a SBN crystal. Due to the inherent energy flow in the anisotropic medium, the nondiffractive beams self-split in two secondary lobes. To enhance the degree of splitting, we use {\textit spatial phase modulation}. Applications of these phenomena to inducing
chiral waveguides and generating even-order vortices are briefly discussed too.

\section{The theoretical model and experimental setup}

\noindent We consider the paraxial light propagation, using the
nondiffractive Lommel-Gaussian beam (NLGB) as the input, which is the
product of the Lommel function and the Gaussian factor \cite{38}:

\begin{equation}
\psi _{\mathrm{NLGB}}(r,\phi ,z=0)=\mathrm{Ga}(r)d^{-n}\mathrm{U}_{\mathrm{n}%
}\left( dk_{t}r\exp (i\phi ),k_{t}r\right) ,  \label{eq:1}
\end{equation}%
where ($r$,$\phi $) and $z$ are the polar coordinates and propagation
distance, Ga$(r)=\exp (-r^{2}/w_{0}^{2})$ is the Gaussian factor with width $%
w_{0}$, $k=2\pi /\lambda $ is the wavenumber ($\lambda $ is the wavelength),
$k_{t}=2\pi /a$ ($a$ is a real constant) is the transverse component of the
wave vector, controlling the scale of the beam, integer $n$ is the quantum
numner of the orbital angular momentum (OAM), and $d$ is the shape factor
taking values $0<|d|<1$. Further, $\mathrm{U}_{n}$ is the Lommel function
defined as:

\begin{equation}
\mathrm{U}_{\mathrm{n}}(s,t)=\sum_{p=0}^{\infty }(-1)^{p}\left( \frac{s}{t}%
\right) ^{n+2p}J_{n+2p}(t),  \label{eq:2}
\end{equation}%
where $J_{n+2p}$ is the Bessel function of the first kind \cite{39}.
Because, as seen in Eq. (\ref{eq:2}), the Lommel function is a superposition
of the Bessel functions of different orders, one may expect that the so
constructed NLGB may feature some properties of Bessel beams, including the
nondiffractive propagation and carrying the OAM, as shown
below. Next, we define the modified Lommel-Gaussian beam (MLGB), by applying
a piecewise-constant modulation profile to it:

\begin{equation}
\psi _{\mathrm{MLGB}}=\left\{
\begin{array}{l}
\psi _{\mathrm{NLGB}},|x|>x_{s}, \\
-\psi _{\mathrm{NLGB}},|x|<x_{s}%
\end{array}%
\right. ,  \label{eq:3}
\end{equation}%
where $x_{s}$ is the splitting scale.

To investigate the propagation of the NLGBs and MLGBs in the nonlinear
medium, we solve the nonlinear Schr\"{o}dinger equation with the input given
by Eqs. (\ref{eq:1}) or (\ref{eq:3}) \cite{40}:

\begin{equation}
i\frac{\partial }{\partial z}\psi (x,y,z)+\frac{1}{2k_{z}}\left[ \nabla
_{\perp }^{2}+V(I)\right] \psi (x,y,z)=0,  \label{eq:4}
\end{equation}%
where $\nabla _{\perp }^{2}$ is the paraxial-diffraction (transverse
Laplace) operator, and $V(I)=-k_{z}^{2}n_{e}^{2}r_{33}E_{\mathrm{sc}}(I)$ is
the effective nonlinear potential, $k_{z}$ represents the longitudinal
wavenumber, $n_{e}$ and $r_{33}$ are, respectively, the extraordinary bulk
refractive index and electro-optic coefficient ($n_{e}=2.2817$ and $%
r_{33}=250$ pm$\cdot $V$^{-1}$ for the SBN crystal), and $I=|\psi |^{2}$ is
the intensity of the input field. Space-charge carriers excited by the input
light field redistribute inside the crystal and form the space-charge
electric field, $E_{\mathrm{sc}}(I)=-\partial \varphi _{\mathrm{sc}%
}(I)/\partial x$. Due to the photorefractive effect, the corresponding
potential $\varphi _{\mathrm{sc}}(I)$ is established across the intensity
profile, as determined by equation \cite{41}

\begin{equation}
\nabla _{\perp }^{2}\varphi _{\mathrm{sc}}+\nabla _{\perp }\ln (1+I)\cdot \nabla
_{\perp }\varphi _{\mathrm{sc}}=E_{\mathrm{ext}}\frac{\partial }{\partial x}%
\ln (1+I)+\frac{k_{B}T}{e}\left[ \nabla _{\perp }^{2}\ln (1+I)+\left( \nabla
_{\perp }\ln (1+I)\right) ^{2}\right] ,  \label{eq:5}
\end{equation}%
where $k_{B}$ is the Boltzmann constant, $T$ the temperature, and $e$ the
elementary charge. In this equation, intensity $I$ is scaled by the dark
intensity $I_{d}$, which arises from the thermal excitation of electrons,
and can be measured experimentally. The effect of the dark intensity on the
beam dynamics was discussed in Ref. \cite{41}. 

\begin{figure}[htbp]
\centering
\includegraphics[width=1\linewidth]{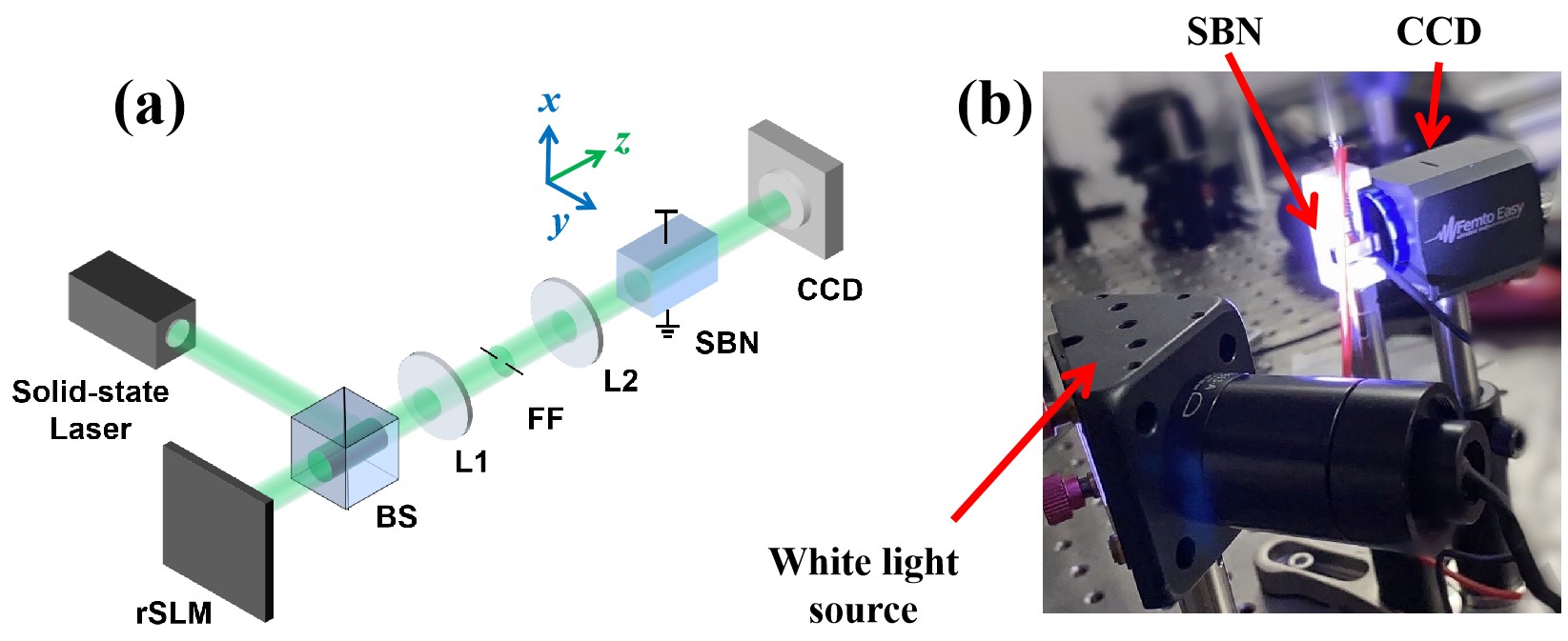}
\caption{The experimental setups for
the
investigation
of
the NLGB
and MLGB propagation in the SBN crystal.
(a)
The schematic of
the
writing
and detecting setups: BS -- beam
splitter; rSLM -- reflective phase-only spatial light modulator; L$1$ and L$2
$ -- lenses with focal length $f=300$ mm; FF -- Fourier filter; CCD --
charge-coupled device. (b) The snapshot of the
erasing
setup.}
\label{fig:1}
\end{figure}

Figure \ref{fig:1}(a) displays the writing and erasing setups used for the
investigation of the
NLGB and the MLGB' propagation in the
SBN crystal. Emitted by a solid-state laser, the $532$ nm
Gaussian beam is divided by the beam splitter in two. One of them
illuminates the reflective spatial light modulator (Santec SLM-200, 1900$%
\times $1200 pixels), onto which a computer-generated hologram, containing
the phase and amplitude information of the lattice-writing beam, is loaded.
Then the extraordinarily polarized beam passes through the $4f$ system
(incorporating two lenses with the focal distance $f=300$ mm and a Fourier
filter) and the photorefractive Cerium-doped SBN crystal (SBN: $61$, $%
5\times 5\times 15$ mm$^{3}$). To enhance the light-matter interaction, the
crystal is externally biased by electric field $E_{\mathrm{ext}}=1600$ V$%
\cdot $cm$^{-1}$ along the optical $c$ axis (i.e., along $x$) \cite{26}.
Eventually, the beam is recorded by the CCD camera. Due to
the memory effect of the refractive-index change, the written photonic
structure has to be erased by white-light illumination. The setup is shown
in Fig. \ref{fig:1}(b). In this setup, the CCD camera monitors the erasing process via
the output intensity pattern.

\section{\label{sec:level3}Results and discussions}

\begin{figure}[tbph]
\centering
\includegraphics[width=1\linewidth]{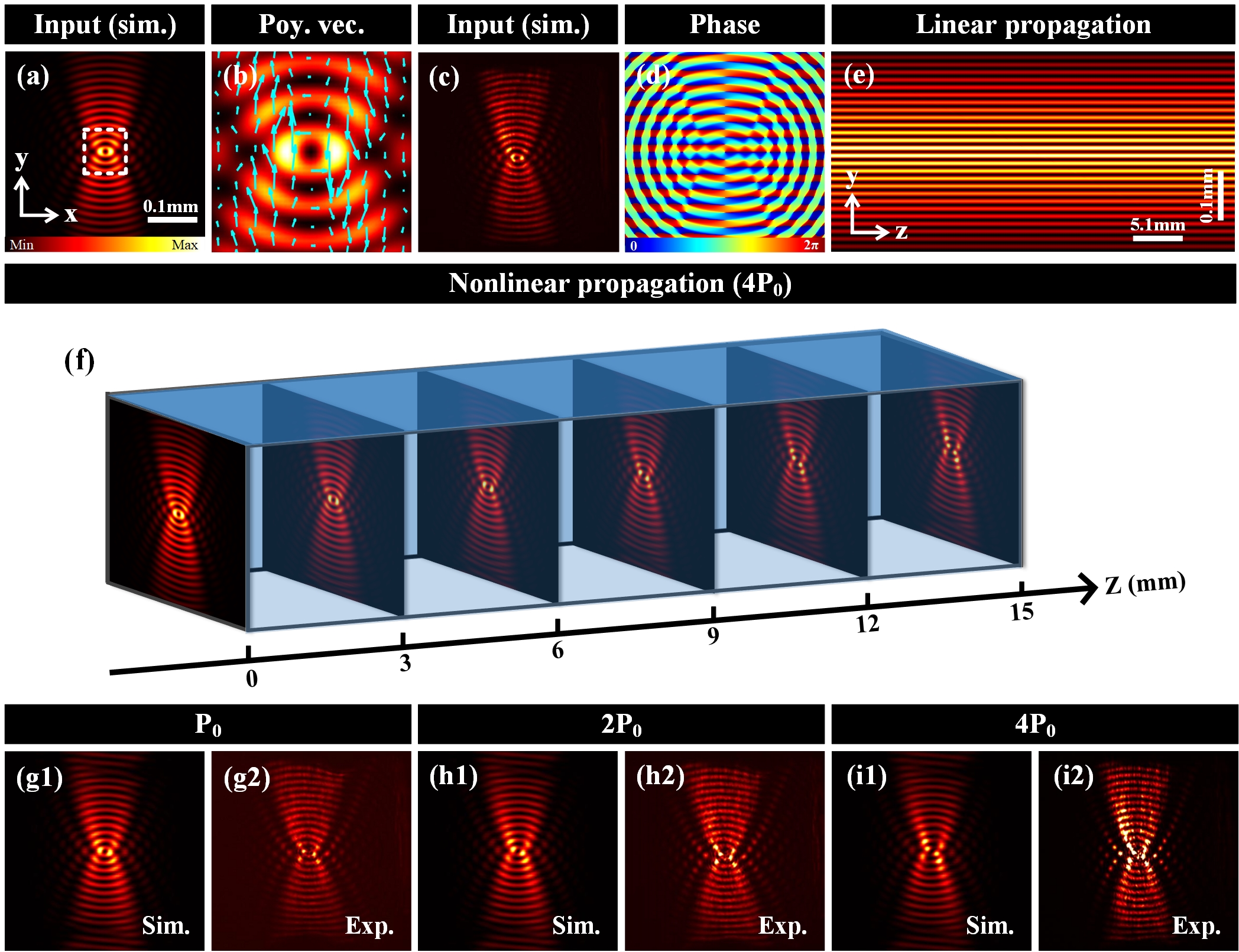}
\caption{The illustration of the NLGB propagation in the free space and in
the photorefractive SBN crystal. (a),
(b)
and
(d)
The
intensity distribution, Poynting vectors, and
phase
distribution
at the
front
face of the crystal ($z=0$), used in
the
simulations.
(c)
The experimental
result corresponds to (a). (e)
Side
view of the
simulated
propagation in the
free space. (f) Cross
sections
of the
simulated NLGB
propagation in the
nonlinear medium with
power
$P=4P_{0}$.
(g1), (h1) and
(i1) are the
simulated output at the
back face
of the
crystal, with powers
$P=P_{0}$,
$P=2P_{0}$, and
$P=4P_{0}$,
respectively,
(g2), (h2) and (i2) being
the
corresponding
experimental
snapshots. The
parameters are
$P_{0}=20.3 
 \mathrm{\mu}$W, $%
d=0.9i$, $\protect\lambda =532 $ nm, $w_{0}=0.25$ mm, $a=0.03$ mm, and $n=1$%
. }
\label{fig:2}
\end{figure}

\noindent First, we simulate the propagation of the NLGB in the SBN crystal
with parameters $d=0.9i$, $\lambda =532$ nm, $w_{0}=0.25$ mm, $a=0.03$ mm
and $n=1 $,
under the injection power $P=4P_{0}$, $P_{0}=20.3 \mathrm{\mu}$W. The intensity and phase distributions in the input plane and the side
view of the simulated propagation in the free space are displayed in Figs. \ref{fig:2}(a), \ref{fig:2}(d) and \ref{fig:2}(e), respectively.
To quantify the energy flux, we calculate the time-averaged Poynting vector
\cite{42}:

\begin{equation}
\langle \vec{S}\rangle =\frac{c_{0}}{4\pi }\langle \vec{E}\times \vec{B}%
\rangle =\frac{c_{0}}{8\pi }\left[ i\omega \left( \psi \nabla _{\perp }\psi
^{\ast }-\psi ^{\ast }\nabla _{\perp }\psi \right) +2\omega k|\psi |^{2}%
\overrightarrow{e_{z}}\right] ,  \label{eq:6}
\end{equation}%
where $c_{0}$ is the light speed in vacuum, $\vec{E}$ and $\vec{B}$ are the
electric and the magnetic fields, and $\omega $ is the optical frequency.
The vectors are indicated by blue arrows in Fig. \ref{fig:2}(b). Note that
nondiffractive beams with an invariant intensity distribution carry inherent
energy flow. However, it is invisible in isotropic media \cite{25}, while
the situation is different in the SBN crystal with the anisotropic
structure. To present the propagation in the crystal, in Fig. \ref{fig:2}(f)
we display simulated cross sections inside the crystal. The inherent energy
flow manifests itself, due to the nonlinearity, at the propagation distance $%
\simeq 9$ mm. The explanation is that, under the action of the
photorefractive effect, electrons excited by the input optical beam drift
transversely to the external field and accumulate, resulting in an increase
of the refractive index \cite{43}. The so built potential hinders the
transverse flux and the energy concentrates in high-intensity spots. On the
other hand, the energy is guided towards the direction perpendicular to the
electric field, where the space-charge potential is relatively weak,
stretching high-intensity spots into two high-energy stripes, as seen in the
snapshot at $15$ mm in Fig. \ref{fig:2}(f).
Our experimental
result, displayed in Fig. \ref{fig:2}(i2) at the back face of the SBN crystal, agrees
well with the theoretical prediction. In addition, we also investigate the
results under different nonlinearity strengths, controlled by the input power.
The simulation and experiments with respect to the intensity distributions
are shown in Figs. \ref{fig:2}(g) and \ref{fig:2}(h). It is seen that the
nonlinear strength determines the formation of high-intensity spots and
high-energy stripes.

\begin{figure}[tbph]
\centering
\includegraphics[width=0.95\linewidth]{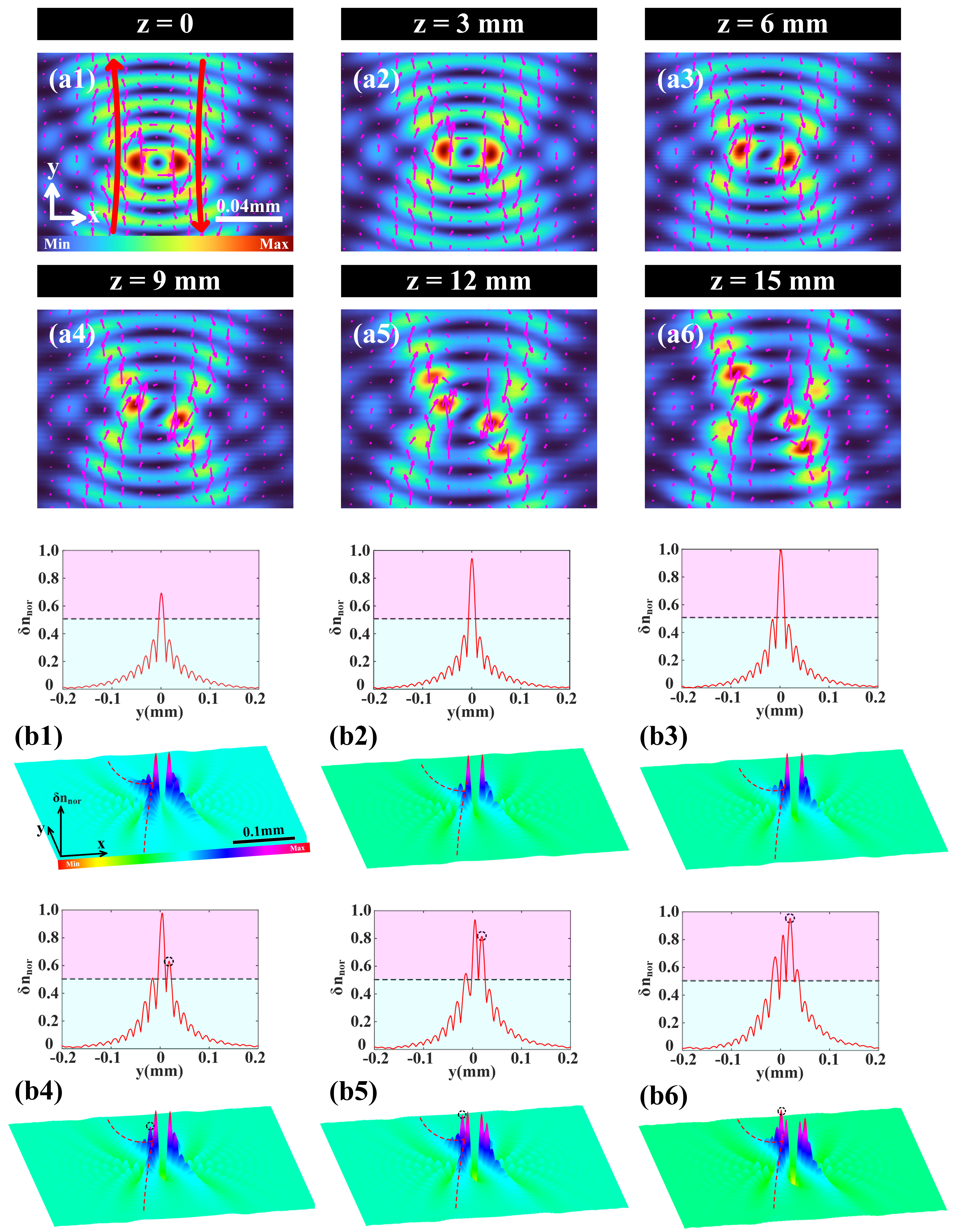}
\caption{The explanation of the self-splitting. (a1)-(a6) The energy density
of the NLGB and Poynting-vector patterns for different propagation
distances. (b1)-(b6) Upper rows: The normalized refractive-index modulation
along the left side of the main lobe, marked by the red dashed line in the
lower rows (the picture is normalized by the maximum value of the
refractive-index change, $\protect\delta n_{\max }=2.29\times 10^{-4}$).
Pink and blue shaded regions designate the normalized values which are
higher and lower than $0.5$, respectively, separated by the black dashed
line. Lower rows: the three-dimensional visualization of the refractive
index modulation, corresponding to (a1)-(a6). }
\label{fig:3}
\end{figure}

The numerical and experimental findings reported in Fig. \ref{fig:2} suggest
that the NLGB's propagation in the photorefractive crystals leads to
self-splitting. To elucidate the physical mechanism of this effect, we
analyze the energy flow, energy density and refractive index modulation, as
shown in Fig. \ref{fig:3} at different values of the propagation distance.
According to the Poynting vectors marked by carmine arrows in Fig. \ref%
{fig:3}(a1), the internal energy flow is predominantly concentrated on both
left and right sides of the beam, being partly distributed transversely, as
indicated by the concentration and length of the arrows. Notably, the direction of the Poynting vectors in
the left part is down-up, while in the right part it is inverse.

Furthermore, taking into regard the nonlinearly-induced
potential redistribution, we present the respective modulation of the local
refractive index, calculated according to Refs. \cite{44,45}:

\begin{equation}
 \delta n(I)=-\frac{1}{2} n_e^3 r_{33} E_{\mathrm{sc}}(I). 
\label{eq:7}
\end{equation}

\noindent The NLGB propagation demonstrates positive feedback in the
focusing nonlinear medium. In upper rows of Figs. \ref{fig:3}(b1)-\ref{fig:3}%
(b6), we show the refractive-index modulation, normalized by the maximum
value of the refractive-index change, $\delta n_{\max }=2.29\times 10^{-4}$,
in the course of the propagation, along left lobes (marked by red dashed
lines in the bottom rows) of the NLGB. To quantitatively analyze the
nonlinearity, we set value $0.5$ as a criterion for it, as indicated by the
black dashed line, with the pink and blue regions above and beneath it
representing relatively strong and weak modulations, respectively. Indeed,
the high-energy central rings and side lobes induce a local increase of the
refractive index, indicated by dark-blue and purple patterns in the bottom
rows of Figs. \ref{fig:3}(b1)-\ref{fig:3}(b6). In the course of the first $6$
mm of the propagation, the central and side peaks rise in the upper rows of
Figs. \ref{fig:3}(b1)-\ref{fig:3}(b3). However, only the central peak
exceeds $0.5$ (from about $0.68$ to $1.0$), indicating the relatively strong
refractive-index modulation. As mentioned above, the self-focusing
nonlinearity drives the positive feedback. The energy concentration results
in a higher local refractive index, followed by accumulating still more
energy at the same spot. The feedback is further observed in the upper row
of Figs. \ref{fig:3}(b4)-\ref{fig:3}(b6). The right-side peak, indicated by
the black dashed circle, exceeds the $0.5$ criterion in Fig. \ref{fig:3}(b4)
and keeps growing higher. By calculating the peaks in the region of the
relatively strong modulation, we find that, in the framework of the
displayed picture,, the number of peaks increases from $1$ to $4$, which is
a cogent demonstration of the positive feedback mechanism.

\begin{figure}[tbph]
\centering
\includegraphics[width=1\linewidth]{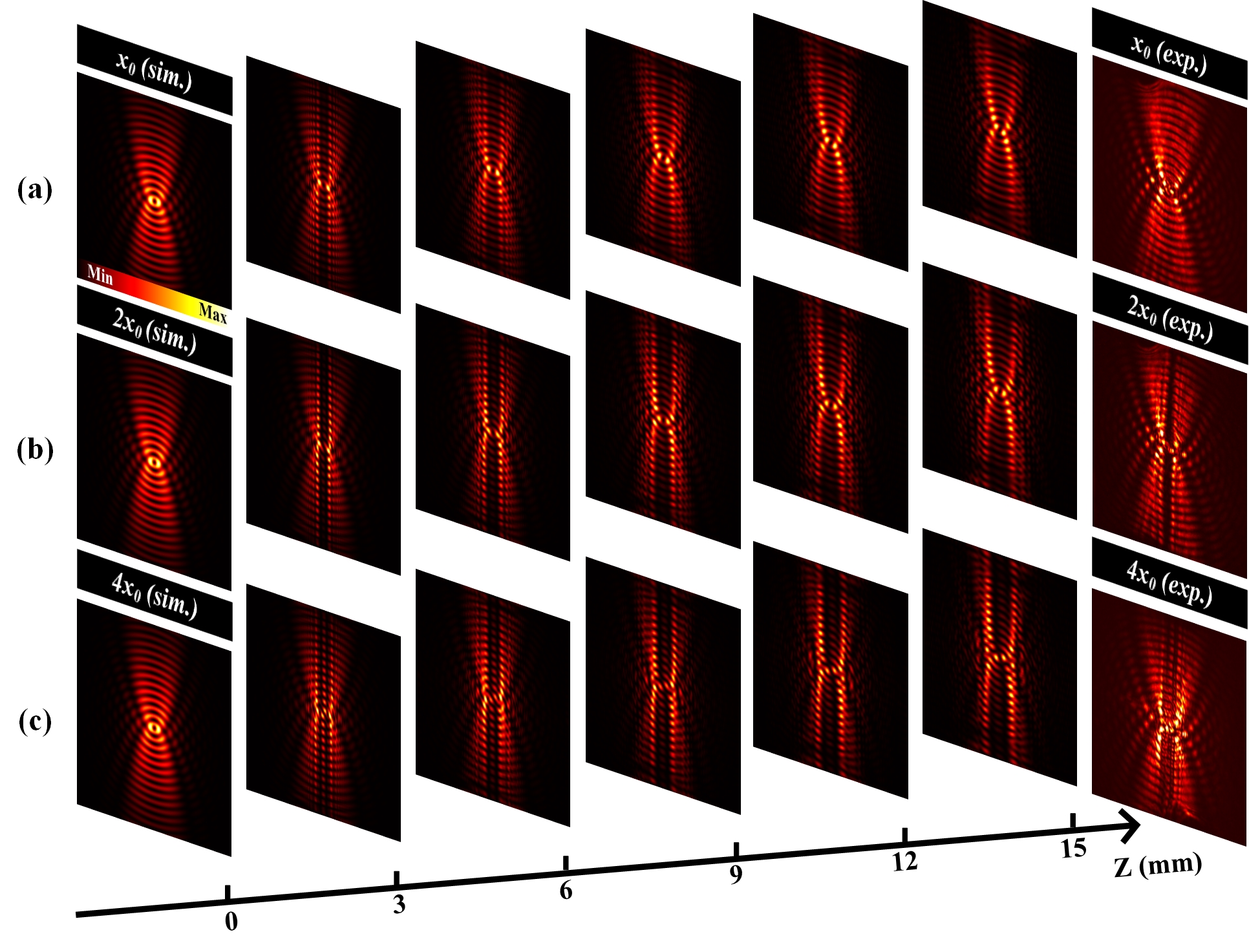}
\caption{Cross sections of MLGBs produced by simulations of their
propagation in the nonlinear medium, with different values of splitting
parameter $x_{s}$ and fixed $x_{0}=0.002$ mm. (a) $x_{s}=x_{0}$; (b) $%
x_{s}=2x_{0}$; (c) $x_{s}=4x_{0}$. The cross sections experimentally
observed at the output face of the crystal are displayed in the far right
column.}
\label{fig:4}
\end{figure}

While direct experimental observation of internal profiles of the
propagating beams is unfeasible, numerical simulations make it possible to
monitor the dynamics inside the SBN crystal. Accordingly, in Fig. \ref{fig:4}%
, we display the simulated evolution of the MLGB cross sections for
different values of the splitting parameter $x_{0}$ and a fixed input power, %
$P=4P_{0}$. Due to the applied additional %
spatial phase modulation, the MLGB splits in the transverse
direction, from the middle towards both edges. The nonlinear dynamics is
essentially the same as explored above. However, the comparison of Figs. \ref%
{fig:4} and \ref{fig:2}(f) demonstrates that the splitting degree rises
under the action of the spatial phase modulation added to
the NLGB. The separation between the two high-energy stripes keeps
increasing as $x_{0}$ rises and the stripes shift sideways. Our experimental
results presented in the far right column of Fig. \ref{fig:4} verify the
prediction.

\begin{figure}[htbp]
\centering
\includegraphics[width=1\linewidth]{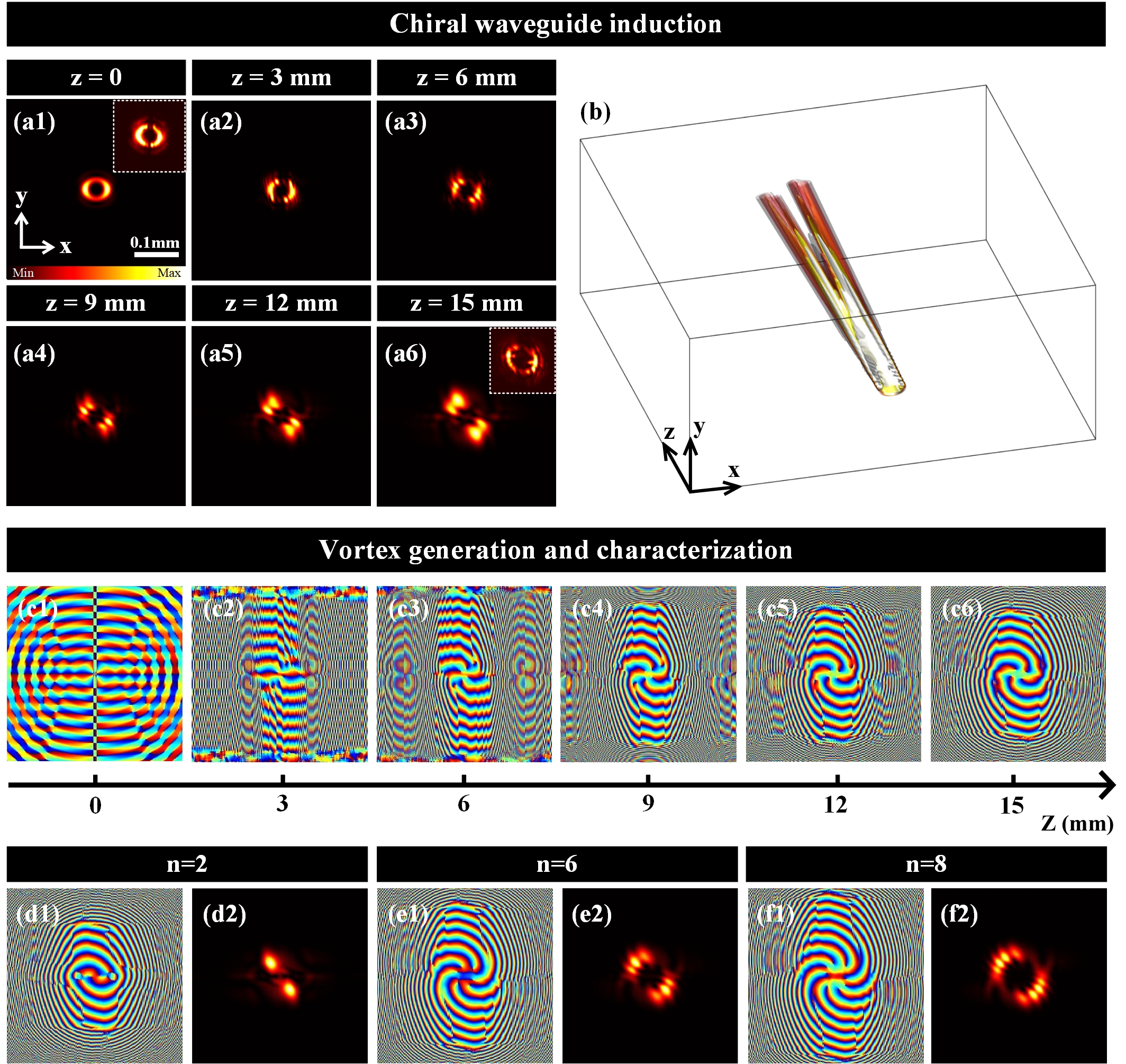}
\caption{The demonstration of the
nonlinearity-induced
chiral waveguide and vortex generation. For the
realization of the chiral
waveguide induction, the input MLGB is taken as
in Fig. \protect\ref{fig:4}(b), but with $n=4$ in Eq. (\protect\ref{eq:1}).
The MLGB with pruned side lobes splits and rotates in the course of the
propagation. Panels (a1)-(a6) display cross sections at different distances,
as produced by the simulations. Insets in (a1) and (a6) show the
corresponding experimental snapshots. (b) The simulated three-dimensional
envelope of the self-splitting beam. Parameters of the MLGBs are $%
x_{s}=2x_{0}$, $w_{0}=a=30$ $\mathrm{\protect\mu }$m and $n=4$, see Eq. (%
\protect\ref{eq:1}). The bottom half of the figure
demonstrates the generation
and characterization of the vortex under the
action of the nonlinearity. (c1)-(c6) The
vortex formation with OAM order
(topological charge) $= 4$. (d1)-(f2)
Simulated vortex phase and intensity
distributions with topological
charges $2$, $6$ and $8$, respectively.}
\label{fig:5}
\end{figure}

It is relevant to consider potential applications of the self-splitting
effect reported in this work. The SBN crystal, due to its unique ability to
keep the memory of the refractive-index changes and the easy-to-erase
property at relatively low writing laser powers \cite{46}, is an ideal
platform for writing flexible and reconfigurable waveguides,
which
was reported in the investigation of spatial rotating solitons in Ref.
\cite{47}. To develop this possibility, we suppress the side lobes of the
MLGB in Fig. \ref{fig:4}(b) and set $n=4$ in Eq. (\ref{eq:1}), to retain the
input's central ring. As shown in Figs. \ref{fig:5}(a1)-\ref{fig:5}(a6), the
OAM-carrying MLGB initially undergoes splitting, with separating
energy-density maxima. Then, two emerging main lobes rotate clockwise,
realizing an effectively chiral waveguide. Further, after passing %
about $9$ mm, the MLGB loses its vorticity, demonstrating
solely diffraction with increasing separation between the main lobes. To
provide an explanation of this behavior, we recall that, according to the
Noether's theorem \cite{48,49}, the angular momentum is not conserved if the
system's rotational symmetry is broken
(of course, the full
angular momentum of the optical field and underlying material medium is
conserved). In the underlying Eq. (\ref{eq:5}), the symmetry is obviously
broken by the anisotropic term with $\partial /\partial x$. Indeed, the
OAM-carrying MLGB rotates by a small angle in Figs. \ref{fig:5}(a1)-\ref%
{fig:5}(a3) and subsequently loses the vorticity in Figs. \ref{fig:5}(a4)-%
\ref{fig:5}(a6). This argument explains the partial rotation of the incident
beams under the action of the nonlinearity,
resembling the effect of the optical field modulation
in the free space \cite{50,51}. Alternatively, chiral waveguides may serve
as an optical circular-dichroism device \cite{52}.
By simulating the phase evolution of the above-mentioned
MLGB inside the SBN crystal, due to the inherent phase singularities defined
by the OAM quantum number $n$, we notice the formation of a vortex with order
4, in Fig. \ref{fig:5}(c6), which matches the number of side lobes at the
output face of the SBN crystal in Fig. \ref{fig:5}(a6). Further, we
simulate the MLGB with $n=2, 6, 8$ in Figs. \ref{fig:5}(d1)-(f2),
and conclude that the matching of side lobes and phase singularities is
a result of imprinting the OAM vorticity (topological charge) from the phase pattern onto the density
one. In other words, by counting the number of side lobes in the output, one can readily
restore the topological charge carried by beam. Due to the degeneracy of
adjacent odd and even topological charges, we here consider only vortices of even orders.
These results can be used for the design of schemes for nonlinear coding
and decoding in optical communications, cf. Refs. \cite{53,54}.

\section{\label{sec:level4}Conclusion}

\noindent In this work we have investigated, theoretically and
experimentally, the propagation of nondiffractive beams in the
photorefractive medium. The energy flow is visualized by the Poynting-vector
field. The light-matter interaction gives rise to the self-splitting of the
beams into separating secondary ones. The splitting degree is effectively
controlled by means of the phase-pre-modulation method. We
have also considered potential applications of the self-splitting for the
design of an effectively chiral waveguide, using the memory feature of the
photorefractive crystal.
The nonlinear generation and characterization of
vortices with even topological charges via output intensity pattern are considered too.

The results are reported in this paper for the self-focusing photorefractive
nonlinearity. We have also performed experiments and simulations for the
system with the self-defocusing sign. These results are not presented here,
as they do not exhibit remarkable effects -- in particular, no splitting is
induced by the self-defocusing.

The experimental and theoretical work can be developed further. In
particular, it may be relevant to consider interactions of two or several
copropagating nondiffractive beams, including a possibility of formation of
their (quasi-) bound states.

\section*{Acknowledgement}

\noindent National Natural Science Foundation of China (12174122 and 11775083);
Guangdong provincial Natural Science Foundation of China (Grant No.
2022A1515011482); Science and Technology Program of Guangzhou (No.
2019050001); Program of innovation and entrepreneurship for undergraduates;
special funds for the cultivation of Guangdong college students scientific
and technological innovation (Climbing Program special funds)
(pdjh2022a0129, pdjh2023a0136); the Extracurricular Scientific Program of
School of Information and Optoelectronic Science and Engineering, South
China Normal University (22GDKB02); Israel Science Foundation (grant No.
1695/22).

\section*{Disclosures}

\noindent The authors declare no conflicts of interest.

\section*{Data availability}
\noindent Data underlying the results presented in this paper are not publicly available at this time but may be obtained from the authors upon reasonable request.

\bibliography{ref}

\end{document}